\documentclass{sigchi-ext}
\usepackage[T1]{fontenc}
\usepackage{textcomp}
\usepackage[scaled=.92]{helvet} % for proper fonts
\usepackage{graphicx} % for EPS use the graphics package instead
\usepackage{balance}  % for useful for balancing the last columns
\usepackage{booktabs} % for pretty table rules
\usepackage{ccicons}  % for Creative Commons citation icons
\usepackage{ragged2e} % for tighter hyphenation
\usepackage{comment}

\usepackage{marginnote} 
\usepackage[utf8]{inputenc} % for a UTF8 editor only

\copyrightinfo{Permission to make digital or hard copies of part or all of this work for personal or classroom use is granted without fee provided that copies are not made or distributed for profit or commercial advantage and that copies bear this notice and the full citation on the first page. Copyrights for third-party components of this work must be honored. For all other uses, contact the Owner/Author.\\
Copyright is held by the owner/author(s).\\
{\emph{MUM '18}}, November 25-28, 2018, Cairo, Egypt.\\
ACM 978-1-4503-6594-9/18/11. \\
https://doi.org/10.1145/3282894.3289740}

\def\plaintitle{Interactive Narrative in Virtual Reality} \def\plainauthor{Gilad Ostrin, Jeremy Frey, Jessica R. Cauchard}

\def\plainkeywords{Interactive Fiction; Virtual Reality; Storytelling; Visualization}

\title{Interactive Narrative in Virtual Reality}
\numberofauthors{3}
\author{%
  \alignauthor{%
    \textbf{Gilad Ostrin}\\
    \affaddr{Interdisciplinary Center Herzliya} \\
    \affaddr{Herzliya, Israel} \\
    \email{gilad.ostrin@post.idc.ac.il } }
    \vfil \alignauthor{%
    \textbf{J\'er\'emy Frey}\\
    \affaddr{Interdisciplinary Center Herzliya}\\
    \affaddr{Herzliya, Israel}\\
    \email{jfrey@ullo.fr}}
    \vfil \alignauthor{%
    \textbf{Jessica R. Cauchard}\\
    \affaddr{Interdisciplinary Center Herzliya}\\
    \affaddr{Herzliya, Israel}\\
    \email{jcauchard@acm.org} \\}
   }

\definecolor{linkColor}{RGB}{6,125,233}
\hypersetup{%draft,
  pdftitle={\plaintitle},
  pdfauthor={\plainauthor},
  pdfkeywords={\plainkeywords},
  bookmarksnumbered,
  pdfstartview={FitH},
  colorlinks,
  citecolor=black,
  filecolor=black,
  linkcolor=black,
  urlcolor=linkColor,
  breaklinks=true,
}

\begin{document}

\maketitle

\begin{abstract}
Interactive fiction is a literary genre that is rapidly gaining popularity. In this genre, readers are able to explicitly take actions in order to guide the course of the story. With the recent popularity of narrative focused games, we propose to design and develop an interactive narrative tool for content creators. In this extended abstract, we show how we leverage this interactive medium to present a tool for interactive storytelling in virtual reality. Using a simple markup language, content creators and researchers are now able to create interactive narratives in a virtual reality environment. We further discuss the potential future directions for a virtual reality storytelling engine.
\end{abstract}





\section{Introduction}
Since the beginning of time, people have been telling stories and created connections through them. Over time, these narrative took different shapes and forms, from oral storytelling traditions to modern technologies such as e-book readers. Given how central storytelling is to our collective psyche and cultures, it comes as no surprise that people constantly look for novel ways to express their imagination. 

In this abstract, we propose a new and interesting paradigm for storytelling in Virtual Reality (VR). We present our interactive narrative concept, its implementation, and describe possible applications for our tool. This tool allows to create a modular and immersive experience in VR, with the goal of adapting a narrative to each reader.

\section{Related Work}

During the 70s, two forms of interactive narratives appeared: Interactive Fiction (IF) and Role Playing Games (RPG). IF encompasses both ``Choose Your Own Adventure'' stories, where readers select branches to advance the story, and ``games'' where readers type text in order to interact with the textual environment \cite{Montfort2003}. 

During this time, pen and paper RPG emerged, with games such as Dungeon's and Dragons \cite{DD}. In such games, players work together to tell a cooperative story, bound by a set of rules and overseen by a "Dungeon Master" to give the story both structure and game elements. This combination of narrative freedom coupled with a grounded rule-base contains useful lessons on how to keep a person engaged and interested in the game.

One of the first IF game, ``Adventure'' \cite{Adventure} had readers interact with a world and characters reminiscent of the fantasy settings of most RPG at the time. In computer-based IF games, the Dungeon Master is replaced by specific algorithms embedded in the software.

Over the last decade IF has gained an increased attention \cite{Montfort2012}, thanks in part to easier writing tools based on markup languages and graphical programming, such as Ink\footnote{\url{https://www.inklestudios.com/ink/}} and Twine\footnote{\url{http://twinery.org/}}. At the same time, pen-and-paper RPG players started to investigate new formats, such as ``Multi-User Dungeons'', a collaborative multi-player version of IF games. Nowadays, one of the most played platform of these games is maintained by Iron Realms\footnote{\url{https://www.ironrealms.com/}}. Another venue consists of online forums such as ``Quests''\footnote{\url{https://forums.sufficientvelocity.com/}}, where one person plays the role of the author, generally telling a story based on a single protagonist. Other players then vote on the next set of actions, followed by the author continuing the story.

IF is a good framework to bring together creativity, flexibility, and adaptation. Yet, it suffers from readers having to explicitly interact with the story. Switching back and forth between reading the narrative and using its interface can impede readers' immersion or the flow of the narration. RPGs possess a collaborative element which can further strengthen the narrative. However, with computerized versions, the role of the Dungeon Master, who interprets the story and serves as the narrator, is often diminished or even non-existent. One notable exception is Neverwinter Nights \cite{NN}, a 3D game that lets one player endorse the role of the Dungeon Master and prepare the narrative. Yet, even there, the limited number of 3D assets can impede creativity.

A main advantage of interactive stories is that readers empathize with protagonists, as if they were part of the story. Adaptive stories have been envisioned as a way to help people connect with unfamiliar situations and characters \cite{Murray1997}. In prior research literature, IFs have been used to study and train empathy. In Cavazza and Charles \cite{Cavazza2016}, brain recordings were used in a virtual environment to detect empathy and alter the narrative. A framework was also described to adapt a 3D narrative using physiology \cite{Gilroy2012}. They used a ``planning domain definition language'', a syntax coming the artificial intelligence domain. These examples are either focused on specific states, or do not intent to bridge the gap with lay writers. 

Our work aims at giving more freedom to content creators, by integrating modern IF tools, and building a platform that can easily be expanded to meet a creator's needs, as discussed in \cite{Frey2016f}. The following section describes our software tool.

\begin{figure}
    \centering
      \includegraphics[width=.8\columnwidth]{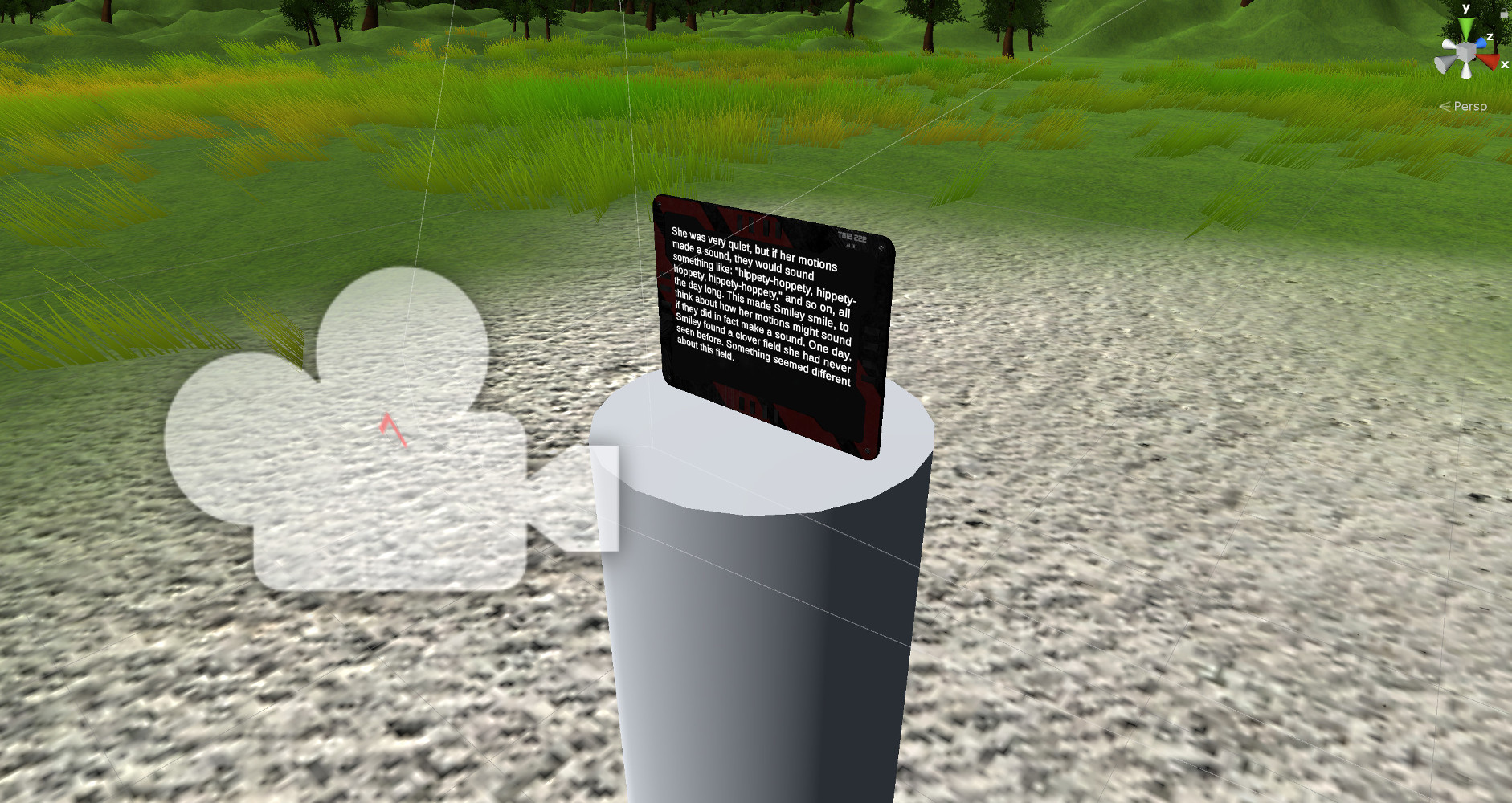}
      \caption{Side view of the VR Scene. The tablet in the center displays the text for the user to read. The thin grey lines represent the user's field of view.}\label{fig:sidepic}
\end{figure}

\section{Interactive Narrative in Virtual Reality}
This section describes our proposed tool for textual storytelling in VR. We first explain the conceptual aspects of the tool, followed by its technical implementation. 

\subsection{Design}
Our tool is implemented in VR, and the story is conveyed through regular text with words. Using text as a medium lets the creator or writer quickly and easily change any element of the story without specific technical knowledge. 

Visually, we propose a simple nature scene in the background. The calming atmosphere of a forest meadow should help offset some of the discomfort that can arise for users new to VR. This meadow is also used as background to the area displaying the text. To increase the users sense of ``presence'' or  ``immersion'', we added a slight grass sway animation that simulates a soft breeze. We did not include sound effects in this first iteration of the tool.

This tool was conceptualized after the ``Choose Your Own Adventure'' paradigm, where the narrative can branch into several directions. In this first iteration, we hide the choice from the user, so that they are unaware that other paths exist. This choice was to simplify the user's experience while allowing for more options in future work. The following section described the implementation stage.

\subsection{Implementation}
The architecture presents two principal components: a Display Engine (DE) and a Selection Engine (SE). The DE handles the content delivery to the user within the experience, and the SE handles the choice of the story path.

For the SE, we wanted to build a separate program that would allow someone outside of the DE to control the flow of the narrative. This 'Director' would make the choices that affected what path the story follows, and thus influence users' experience. We opted for a simple interface with the story options as buttons for this purpose.

We implemented the Display Engine using Unity, as it has the right balance of features and ease of use. The forest meadow scene background was built using Unity features, and its physics engine was used for the movement of the grass. A publicly available tablet mesh was then positioned at the user's eye level to display the story. The text is displayed onto the virtual tablet using Text Mesh Pro Unity library, which allows advanced text features to be used in the stories. This includes high level HTML markup features, such as changing color, position, font, or alignment on the fly. To further support future research, we used the underlying mesh of each word to create a collider surrounding it. This collider can be used in future experiments for eye or head tracking, or to create dynamic ``interactive areas'' inside the story itself.

The SE component contains the interface for a Director, the bridge to the DE, and allows for control of the story. It was built using C\#. The Director's Interface was created using a Windows Form program that displays the current story status, and dynamically creates buttons for the Director to control the story flow. On the Unity side, a C\# script connects to the Director's Interface and follows its commands. Lab Streaming Layer (LSL) was used for the bridge, given its compatibility with both C\# and Unity.
Finally, we use the ``Ink'' scripting language to write the stories. Ink is a language specialized for this purpose, allowing for non-technical users to easily write branching and dynamic stories.

\begin{figure}
\centering
  \includegraphics[width=1\columnwidth]{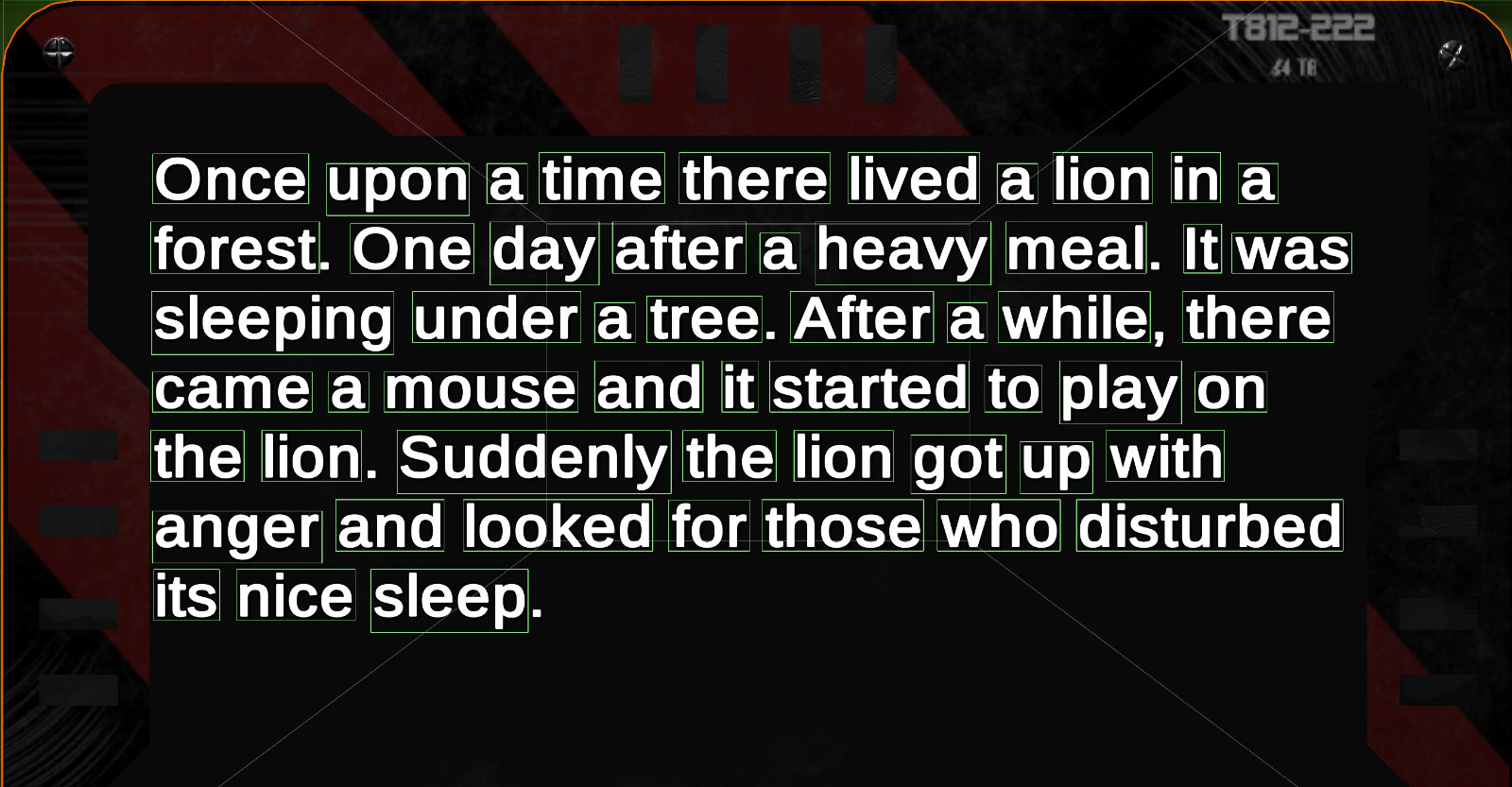}
  \caption{Example demonstrating how the text is displayed in VR.}\label{fig:collider}
\end{figure}

\section{Applications}
There are several avenues possible when considering the applications of this tool and technology, beyond supporting creative writers. One is to use it for research and user studies. For example, a researcher could build various versions of a story and test how the different versions affect people, such as when looking at emotional response or empathy. In this situation, the researcher would be able to easily and quickly build, test, and identify the practical effects of small changes within the story.

Another application relates to tailoring content to fit specific age or demographics. An author could create a story using several levels of language, and dynamically adjust it so that the reader can more easily comprehend the content. A news article could use simpler words for a teenager or a child, while a college student might get a more thorough description. This could allow for the same content to be absorbed better according to the readers' needs.

\section{Evaluation and Preliminary Results}
After having performed several pilot studies, we planned to formally evaluate the performance of this tool in three stages: 1) a technical evaluation, 2) a first user study to see how various people react to different stories, some of which being personally tailored, and 3) a second user study to see how this tool could help support content creators.

The technical evaluation will support evaluating the effectiveness of the tool. This includes reviewing the usability of the tool by non-technical users, checking the viability of the Director component, and testing the Director's LSL extension via an automated remote program. We will measure usability by asking non-technical creative writers to write stories that utilize the branching behavior. A usability questionnaire will be administrated to assess the interface ease of use. The viability of the Director will be tested by performing tests to estimate the minimal power and processing requirements of the system. 
To test the Director's LSL extension, we wrote an automated script that can choose a story path at random for the reader. 

Next, we will write a set of stories with several branching options, such as the ability to choose a male or female protagonist. We would then change the story to accommodate the reader, and investigate if a tailored story can promote empathy or foster playfulness. This result would be compared to a control group who has the story parameters set at random. Should this step be successful, we would then move on to stage 3, where we streamline the tool for use with existing content creators.

\section{Future Work and Conclusions}
There are a number of avenues to explore in future work. We envision reaching person's empathy by tailoring the narrative. We will further explore the possibility of creating an artificial intelligence Director to control the scene. Such directors are common in gaming, and we believe this concept can be adapted to a literary experience.

In conclusion, we have presented a tool and a novel approach for creating interactive narrative fiction in virtual reality. Our tool is a stepping stone to a different means of engaging in storytelling. It allows for content creators with no technical experience to design relatable stories in VR. With its simple concept and design, more advanced versions of this tool can be created for specific applications. 

\balance{} 

\bibliographystyle{SIGCHI-Reference-Format}
\bibliography{sample}

\end{document}